\documentclass[aps,twocolumn,prb,amsfonts,showpacs,superscriptaddress,longbibliography]{revtex4-2}
\usepackage[utf8]{inputenc}
\usepackage{verbatim,epsfig,amsmath,amssymb,bm,epsf,graphicx,psfrag,bbold,amsthm,amsfonts}
\usepackage[colorlinks,linkcolor=blue,anchorcolor=blue,citecolor=blue,urlcolor=blue]{hyperref}
\usepackage[bottom]{footmisc}
\usepackage{hyperref}
\usepackage{cleveref} 
\usepackage{enumitem}
\usepackage{framed}
\usepackage{mathrsfs}
\usepackage{esint}
\setlist{nosep}
\usepackage{placeins}
\usepackage[all]{xy}
\usepackage{color}
\usepackage[utf8]{inputenc}
\usepackage{float}
\usepackage{natbib}
\usepackage{tikz-cd}
\usepackage{verbatim}
\usepackage{leftidx}
\usepackage{physics}
\usepackage{ulem}
\usepackage{gensymb}
\usepackage{soul}
\usepackage{xcolor}
\usepackage{comment}

\usepackage{makecell}

\date{\today}

\begin{document}
\title{Topological phase transitions in twisted bilayer graphene/hBN from interlayer coupling and substrate potentials}

\author{Huiwen Wang}
\affiliation{Centre for Quantum Physics, Key Laboratory of Advanced Optoelectronic Quantum Architecture and Measurement (MOE), School of Physics, Beijing Institute of Technology, Beijing, 100081, China}
\author{Wei Jiang}
\email{wjiang@bit.edu.cn}
\affiliation{Centre for Quantum Physics, Key Laboratory of Advanced Optoelectronic Quantum Architecture and Measurement (MOE), School of Physics, Beijing Institute of Technology, Beijing, 100081, China}
\affiliation{International Center for Quantum Materials, Beijing Institute of Technology, Zhuhai, 519000, China}

\begin{abstract}
Twisted bilayer graphene aligned with hexagonal boron nitride (TBG/hBN) hosts rich topological and correlated quantum phases, such as (fractional) Chern insulators, whose character is dictated by the topology of the moir\'{e} flat band. This topology is highly sensitive to several material parameters in the continuum model, yet a systematic understanding of their combined influence has been lacking. Here, we present a comprehensive study of topological phase transitions in TBG/hBN by varying the interlayer hopping strengths ($w_0, w_1$) and hBN-induced staggered potential, both with and without the hBN moir\'{e} potential. We map out Chern number phase diagrams across a broad, experimentally relevant parameter space, revealing a progressive enrichment of the topological landscape including multiple high-Chern number ($C$ = 3, 4, and 5) states. Each transition is linked to distinct band-inversion mechanisms at generic $C_3$-symmetric k points, high-symmetry momenta, or parabolic touchings, clearly reflecting in the evolution of the Berry curvature. Our results offer theoretical insights that help interpret existing experimental observations, elucidate the mechanisms driving these topological phase transitions and facilitate the exploration of topological states in TBG/hBN and related moir\'{e} systems.
\end{abstract}

\maketitle

\section{Introduction}

Twisted van der Waals materials have emerged as a central platform in modern condensed matter physics, offering unprecedented access to a rich landscape of correlated and topological quantum phases. Key experimental discoveries encompass unconventional superconductivity, correlated insulators, magnetic orders, and topological states such as the quantum anomalous Hall effect (QAHE) and fractional Chern insulators \cite{caoUnconventionalSuperconductivityMagicangle2018, poOriginMottInsulating2018, isobeUnconventionalSuperconductivityDensity2018, wuTheoryPhononMediatedSuperconductivity2018, caoCorrelatedInsulatorBehaviour2018, zhangTwistedBilayerGraphene2019, kangStrongCouplingPhases2019, songDirectVisualizationMagnetic2021, yangMoireMagneticExchange2023, jeongInterplayValleyLayer2024a, songAllMagicAngles2019, songTwistedBilayerGraphene2021, bernevigTwistedBilayerGraphene2021}. In particular, QAHE with tunable Chern numbers has been observed in twisted bilayer graphene (TBG) aligned with hexagonal boron nitride (hBN) \cite{sharpeEmergentFerromagnetismThreequarters2019, stepanovUntyingInsulatingSuperconducting2020}, as well as under applied magnetic fields \cite{wuChernInsulatorsVan2021, dasSymmetrybrokenChernInsulators2021, nuckollsStronglyCorrelatedChern2020, saitoHofstadterSubbandFerromagnetism2021a, parkFlavourHundsCoupling2021a}, while FCIs have been reported in twisted MoTe$_2$ and rhombohedral graphene/hBN stacks \cite{parkObservationFractionallyQuantized2023, luFractionalQuantumAnomalous2024, luExtendedQuantumAnomalous2025, xieTunableFractionalChern2025}. Crucially, these diverse phenomena are intimately linked to the topology of the moir\'{e} flat bands, highlighting the fundamental role that band topology plays in driving the emergent quantum physics of twisted systems.

Theoretical investigation of these systems relies heavily on the continuum model, which effectively captures the long-wavelength physics of interlayer coupling and moir\'{e} periodicity \cite{bistritzerMoireBandsTwisted2011, wuTopologicalInsulatorsTwisted2019}. This approach is complemented by other methods, including tight-binding models \cite{koshinoMaximallyLocalizedWannier2018, suarezmorellFlatBandsSlightly2010, poFaithfulTightbindingModels2019, yuGeneralElectronicStructure2025}, large-scale density functional theory (DFT) \cite{maoTransferLearningRelaxation2024}, and machine-learning-accelerated techniques for handling structural relaxation \cite{liDeeplearningDensityFunctional2022, baoDeepLearningDatabaseDensity2024, liuDPmoireToolConstructing2025}. When an hBN substrate is introduced, the continuum model must account for several key parameters: the interlayer coupling strengths ($w_0-w_1$), a sublattice-dependent staggered potential, and a moir\'{e} potential et al. These parameters are typically fitted from DFT calculations \cite{zhangUniversalMoireModelBuildingMethod2024, liTwistangleDependenceProximity2019}, but their experimental values can vary significantly due to strain, dielectric environment, external electric field, and precise interlayer alignment, making the topology of the flat bands highly sensitive to realistic conditions.

Previous theoretical studies in TBG/hBN system have mapped out topological phase diagrams for specific parameter ranges \cite{PhysRevB.103.075122, bultinckMechanismAnomalousHall2020, maoQuasiperiodicityBandTopology2021, zhangNearlyFlatChern2019}. For instance, displacement-dependent \cite{PhysRevB.103.075122} and staggered-potential-dependent \cite{bultinckMechanismAnomalousHall2020} diagrams have been explored. However, a systematic investigation of topological phase transitions across the combined, physically relevant window of all key parameters remains lacking. To bridge this gap and account for experimental variability, a comprehensive study that tracks the evolution of Chern numbers over a broad parameter space is essential.

In this work, we systematically study the topological phase transitions in hBN-aligned TBG by varying the interlayer hopping strengths ($w_0-w_1$) and the substrate-induced staggered/moir\'{e} potential. We employ the continuum model to compute Chern-number phase diagrams across a wide, experimentally relevant parameter window. In Sec.~\ref{sect2:ctm_model}, we introduce the continuum model and outline our computational methods, including how topological phases can be tuned experimentally. In Sec.~\ref{1},~\ref{2}, and~\ref{3}, we present our numerical results for $w_0-w_1$, $w_0-\Delta_M$, and $\Delta_{M1}-\Delta_{M2}$ topological phase diagrams, respectively, with and without the moir\'{e} potential. For each parameter set, we illustrate the representative phase transitions by tracking the evolution of the band structure and Berry curvature distribution, which reveal band inversions at distinct symmetry points. Notably, we observe the emergence of multiple high-Chern-number phases, with Chern numbers changing by $\pm1, \pm2$, and $\pm3$ across transitions, dictated by the symmetry of the underlying band inversions. We conclude in Sec.~\ref{Summary and discussion} by discussing the experimental relevance of our findings and offering a summary with future perspectives.

\section{model and method}
\label{sect2:ctm_model}

In the TBG/hBN system, two distinct moir\'{e} patterns emerge: one from the graphene-graphene twist (M$_{GG}$) and another from the graphene-hBN stacking (M$_{GBN}$), characterized by twist angles $\theta$ and $\theta^{\prime}$, respectively, as shown in Fig.~\ref{fig:w0-w1 without moire}(a). Here, $a_G$ and $a_{BN}$ denote the lattice constants of graphene and hBN, respectively. The electronic structures of this combined system is accurately captured by a continuum model~\cite{bistritzerMoireBandsTwisted2011}.
The total Hamiltonian is separated into a term describing the twisted bilayer graphene ($H_G$) and a term capturing the effect of the hBN ($V_{BN}$)
\begin{equation}
    H_{G/BN} = H_G + V_{BN}.
\end{equation}
Further details of the model are provided in Appendix~\ref{appendix:BM model}.

The low-energy states of graphene near the Dirac points are dominated by $p_z$ orbitals. Within the $k\cdot p$ framework, the moir\'{e} Hamiltonian for TBG is expressed as
\begin{equation}
    H_G(\boldsymbol{r})=\begin{pmatrix} -\hbar v_f\boldsymbol{q}\cdot\sigma^{\mu} & T(\boldsymbol{r}) \\ T^\dagger(\boldsymbol{r}) & -\hbar v_f\boldsymbol{q}\cdot\sigma^{\mu} \end{pmatrix}
    \label{BM hamiltonian},
\end{equation}
with interlayer coupling matrix:
\begin{align}
    T(\boldsymbol{r})=&\begin{pmatrix} w_0 & w_1 \\ w_1 & w_0 \end{pmatrix} + e^{i\mu\boldsymbol{b}_{m2} \cdot \boldsymbol{r}} \begin{pmatrix} w_0 e^{-i\mu \phi} & w_1 \\ w_1 e^{i\mu \phi} & w_0 e^{-i\mu \phi} \end{pmatrix} \notag \\
    &+ e^{-i\mu{\boldsymbol{b}_{m1}} \cdot \boldsymbol{r}} \begin{pmatrix} w_0 e^{i\mu\phi} & w_1 \\ w_1 e^{-i\mu\phi} & w_0 e^{i\mu\phi} \end{pmatrix},
    \label{interlayer coupling}
\end{align}
where $\sigma^\mu=(\mu\sigma_x,\sigma_y)$ with $\mu=1,-1$ labeling the $K^{\prime}$ and $K$ valleys. $\sigma_x$ and $ \sigma_y$ are Pauli matrices. The model parameters fall into two categories: i) Intrinsic material parameters, determined solely by monolayer graphene: the phase factor $\phi =\frac{2\pi}{3}$ and the Fermi velocity $\hbar v_f=5.25\text{eV}$\AA. ii) Structurally dependent, tunable parameters: the momentum $\boldsymbol{q}$ determined by the distance between Dirac points of two layers, the moir\'{e} reciprocal lattice vectors $\boldsymbol{b}_{m1}, \boldsymbol{b}_{m2}$ (where $\boldsymbol{b}_{m1}, \boldsymbol{b}_{m2}$ are twist-angle dependent), and the interlayer hopping amplitudes $w_0$ (AA/BB) and $w_1$ (AB/BA) that vary with the interlayer distance.

Note that for a general twist angle, the two moir\'{e} patterns are incommensurate and form a longer ‘supermoir\'{e}’ scale, where lattice relaxation can lead to domain formation and qualitatively affect the electronic properties~\cite{nakatsujiMoireBandEngineering2025,zhuModelingMechanicalRelaxation2020,mengCommensurateIncommensurateDouble2023,turkelOrderlyDisorderMagicangle2022,craigLocalAtomicStacking2024,liSymmetryBreakingAnomalous2022,linEnergeticStabilitySpatial2022,shinElectronholeAsymmetryBand2021,nakatsujiMultiscaleLatticeRelaxation2023,parkUnconventionalDomainTessellations2025}. In this work, we focus exclusively on the perfectly commensurate condition where the two patterns are aligned, leaving the supermoir\'{e} regime for future investigation. The influence of the hBN substrate on TBG is captured through two primary effects: a staggered potential ($\Delta_M$) and a moir\'{e} potential arising from the graphene-hBN supercell~\cite{jungInitioTheoryMoire2014}. These are incorporated into the Hamiltonian as
\begin{equation}
    V_{\text{BN}} =  \Delta_M \sigma_z + \sum_{j=1}^{6} V(\boldsymbol{G}^{\prime}_j) e^{i\boldsymbol{G}^{\prime}_j\cdot\boldsymbol{r}},
\end{equation}
where $\sigma_z$ is the corresponding Pauli matrix. The Fourier components of the moir\'{e} potential for the six lowest reciprocal vectors $\boldsymbol{G}^{\prime}_j$ of the twisted graphene-hBN superlattice are given by:
        \begin{align}
        \label{eq:moire potential}
        V(\boldsymbol{G}^{\prime}_1) = V^\dagger(\boldsymbol{G}^{\prime}_4) &= \begin{pmatrix} C_0 + C_z & C_{AB} \\ C_{AB} & C_0 - C_z \end{pmatrix}, \notag \\ 
        V(\boldsymbol{G}_3^{\prime}) = V^\dagger(\boldsymbol{G}^{\prime}_6) &= \begin{pmatrix} C_0 + C_z & e^{-i \frac{2\pi}{3}} C_{AB} \\ e^{i \frac{2\pi}{3}} C_{AB} & C_0 - C_z \end{pmatrix},  \notag \\
        V(\boldsymbol{G}^{\prime}_5) = V^\dagger(\boldsymbol{G}^{\prime}_2) &= \begin{pmatrix} C_0 + C_z & e^{i \frac{2\pi}{3}} C_{AB} \\ e^{-i \frac{2\pi}{3}} C_{AB} & C_0 - C_z \end{pmatrix}.
        \end{align}
\ignorespacesafterend    
Here, we adopt the parameter values $C_0=10.13e^{i(-93.47\degree)}$ meV, $C_z=9.01e^{i(-171.57\degree)}$ meV, and $C_{AB}=11.34e^{i(100.40\degree)}$ meV~\cite{jungMoireBandModel2017}. The staggered potential $\Delta_M$ is set to an experimentally observed value of approximately 15 meV~\cite{huntMassiveDiracFermions2013}. The moir\'{e} potential is significant only within a narrow range of small twist angles, decaying rapidly as the angle increases~\cite{kimAccurateGapDetermination2018}. A pronounced moir\'{e} potential emerges when the graphene-graphene and graphene-hBN moir\'{e} patterns overlap coherently, requiring their reciprocal lattice vectors to be same or related by a 60$\degree$ rotation~\cite{PhysRevB.103.075122,maoQuasiperiodicityBandTopology2021,shinElectronholeAsymmetryBand2021}. The reciprocal lattice vectors are $\textbf{b}_{GG}=\frac{4 \pi}{\sqrt{3} a_G}\left(\sin \theta, 1-\cos \theta\right)$ for the M$_{GG}$ and $\textbf{b}_{GBN}=\left(\frac{4 \pi}{\sqrt{3} a_{B N}} \sin \theta^{\prime}, \frac{4 \pi}{\sqrt{3} a_{B N}} \cos \theta^{\prime}-\frac{4 \pi}{\sqrt{3} a_G}\right)$ for the M$_{GBN}$. When the two moir\'{e} patterns are aligned with a 60$^{\circ}$ rotation, this geometric matching condition imposes a specific relation between the twist angles $\theta$ and $\theta^{\prime}$:
\begin{align}
\label{eq:theta and theta'}
    &\frac{4 \pi}{\sqrt{3} a_G}\left(\sin \theta, 1-\cos \theta\right)R_{60}=  \notag \\
   & \left(\frac{4 \pi}{\sqrt{3} a_{B N}} \sin \theta^{\prime}, \frac{4 \pi}{\sqrt{3} a_{B N}} \cos \theta^{\prime}-\frac{4 \pi}{\sqrt{3} a_G}\right),
\end{align}
where $R_{60}=\begin{pmatrix}
\frac{1}{2}   &-\frac{\sqrt{3} }{2}  \\
  \frac{\sqrt{3} }{2} &\frac{1}{2} 
\end{pmatrix}$ is rotation matrix with a 60$^\circ$ degree. The lattice constant of graphene and hBN are taken as $a_G$ = 2.46 \AA{} and $a_{BN}$ = 2.50 \AA, respectively~\cite{castronetoElectronicPropertiesGraphene2009,westonNativePointDefects2018}. Solving Eq.~\ref{eq:theta and theta'} with these values yields two solutions ($\theta =1.05\degree$, $\theta =0.53\degree$) and ($\theta =1.06\degree$, $\theta ^{\prime}= -0.55\degree$). Throughout this work, we adopt the alignment condition $\theta = 1.05\degree$, $\theta ^{\prime} = 0.53\degree$ for demonstration.

Crucially, all these parameters are experimentally accessible. The interlayer hopping $w_0$ and $w_1$ can be adjusted by modified by varying the interlayer distance (e.g., via pressure). The staggered potential $\Delta_M$ can be tuned by an external perpendicular electric field~\cite{kimAccurateGapDetermination2018, yankowitzDynamicBandstructureTuning2018}, and the moir\'{e} potential can be enhanced by applied pressure~\cite{wangPressureDrivenMoirePotential2025}.
In summary, $w_0, w_1, \Delta_{M}$, and the moir\'{e} potential coefficients $V(\boldsymbol{G_j})$ are key, tunable experimental knobs that significantly influence band topology. In the following section, we systematically analyze how variations in these parameters drive distinct topological phase transitions. 
The methods for calculating the Chern number and Berry curvature are detailed in Appendix~\ref{appendix:numerical calaulation of chern}.

\section{results}

In this section, we explore topological phase transitions in Bernal-stacked TBG/hBN under varying physical conditions, beginning with the influence of interlayer coupling strengths $w_0$-$w_1$ in the absence of a moir\'{e} potential from the hBN substrate. 

\subsection{$w_0$-$w_1$ topological phase diagram}
\label{1}

\begin{figure}[t]
\centering
    \includegraphics[width=\linewidth]{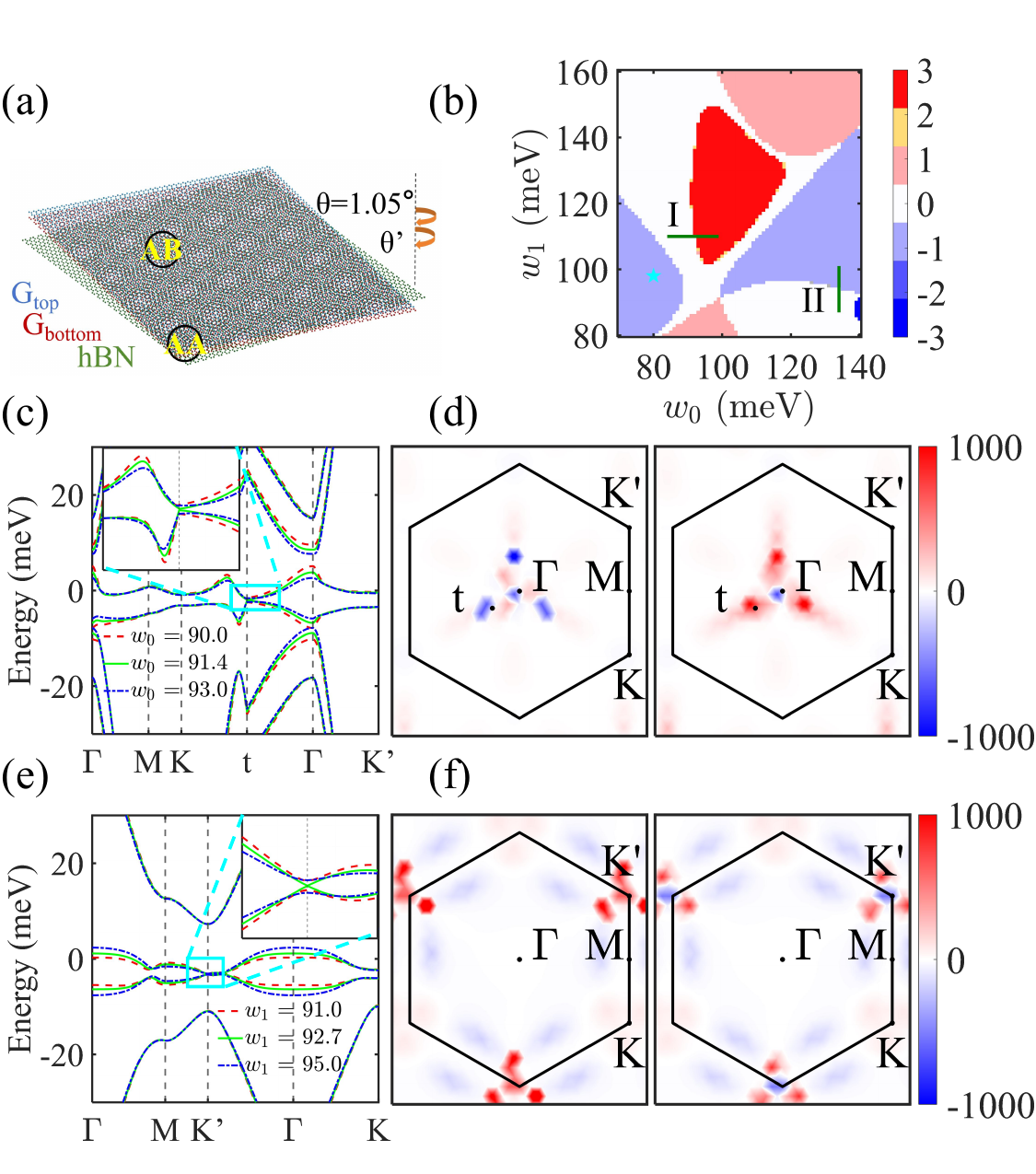}
    \caption{Topological phase transition in TBG/hBN driven by interlayer coupling. (a) Atomic schematic: TBG (blue/red) on a misaligned hBN substrate (green). The twist angle between graphene layers is $\theta$ = $1.05\degree$; the hBN is misaligned at a larger angle $\theta^\prime$. AA and AB stacking regions of TBG are indicated. (b) Chern number phase diagram for the first valence band with $\Delta_{M}$ = 15 meV. The star marks the parameter set from Ref.~\cite{liuOrbitalMagneticStates2021} with $w_0$ = 80 meV, $w_1$ = 98 meV. Two representative topological transitions are labeled I and II. 
    (c) Band structures near the band inversion point for transition I ($w_0$ = 91.4 meV, $w_1$ = 110 meV). (d) Berry curvature before (left: $w_0$ = 91 meV, C = 0) and after (right: $w_0$ = 93 meV, C = 3 ) transition I. (e) Band structures near the inversion point for transition II ($w_0$ = 140 meV, $w_1$ = 92.7 meV). (f) Berry curvature before (left: $w_1$ = 92 meV, C = 0) and after (right: $w_1$ = 94 meV, C = $-1$) transition II.}
    \label{fig:w0-w1 without moire}
\end{figure}

Figure~\ref{fig:w0-w1 without moire}(b) presents the calculated Chern number for the first valence band across the $w_0$-$w_1$ parameter plane, with a fixed staggered potential $\Delta_M=15$ meV. The resulting phase diagram reveals Chern numbers of 0, $\pm$1, $\pm$3, demonstrating a high sensitivity of band topology to the interlayer interaction parameters. For the commonly used values of $w_0$ and $w_1$ obtained from full structural relaxation, we find $C=-1$, consistent with previous theoretical studies~\cite{zhangTwistedBilayerGraphene2019,liuPseudoLandauLevel2019,bultinckMechanismAnomalousHall2020}. For reference, a cyan star marks the parameter set ($w_0$ = 80 meV, $w_1$ = 98 meV, $\Delta_M$ = 15 meV) from Ref.~\cite{liuOrbitalMagneticStates2021}. Notably, as both the $w_0$ and $w_1$ increase, the Chern number can transition to $C= \pm3$. This tunability underscores the TBG/hBN system as a versatile platform for engineering topological states. 

In this phase diagram, we identify two primary phase transitions: transition I ($C = \pm$3 $\leftrightarrow$ 0) and transition II ($C = \pm$1 $\leftrightarrow$ 0). To elucidate their mechanisms, we analyze two representative cases labeled I and II. For transition I, we examine the evolution of the band structure and Berry curvature with varying $w_0$. Figure~\ref{fig:w0-w1 without moire}(c) shows the band structure evolution of TBG/hBN system for $w_1$ = 110 meV and $w_0=90, 91.4, 93$ meV with inset highlighting the middle two flat bands. As $w_0$ increases, the gap between the first conduction and valence bands narrows and closes at a generic point t near the $\Gamma$-K line of the Brillouin zone (BZ) when $w_0=91.4$ meV. This gap closure and reopening correspond to a Chern number change of 3. The associated Berry curvature of the first valence band, shown in Fig.~\ref{fig:w0-w1 without moire}(d) for $w_0=91$ and $93$ meV, reveals a sign reversal at three equivalent t points, consistent with the band inversion picture. We note that due to the finite k-mesh used for Berry curvature calculations, their distributions may appear to lack the $C_3$ symmetry of the system; this symmetry is restored upon using a finer mesh. The emergence of a topological charge of 3 is a direct consequence of threefold band inversions at these $C_3$-symmetric points.

For transition II, we fixed $w_0$ = 140 meV and vary $w_1$. The band structure for $w_1=91, 92.7$, and $95$ meV are shown in Fig.~\ref{fig:w0-w1 without moire}(e). Here, a band inversion occurs at the high-symmetry $K^{\prime}$ point, driving a change in Chern number by $\pm1$, as shown in the inset for the enlarged band structure. The Berry curvature distribution shown in Fig.~\ref{fig:w0-w1 without moire}(f) further confirms this point, where sign change at $K^{\prime}$ point can be clearly seen. The locations of all band inversion points associated with the observed phase transitions in Fig.~\ref{fig:w0-w1 without moire}(b) are cataloged in Appendix~\ref{appendix:B}, Fig.~\ref{fig:direct_gap}.

\begin{figure}[h]
    \centering
    \includegraphics[width=\linewidth]{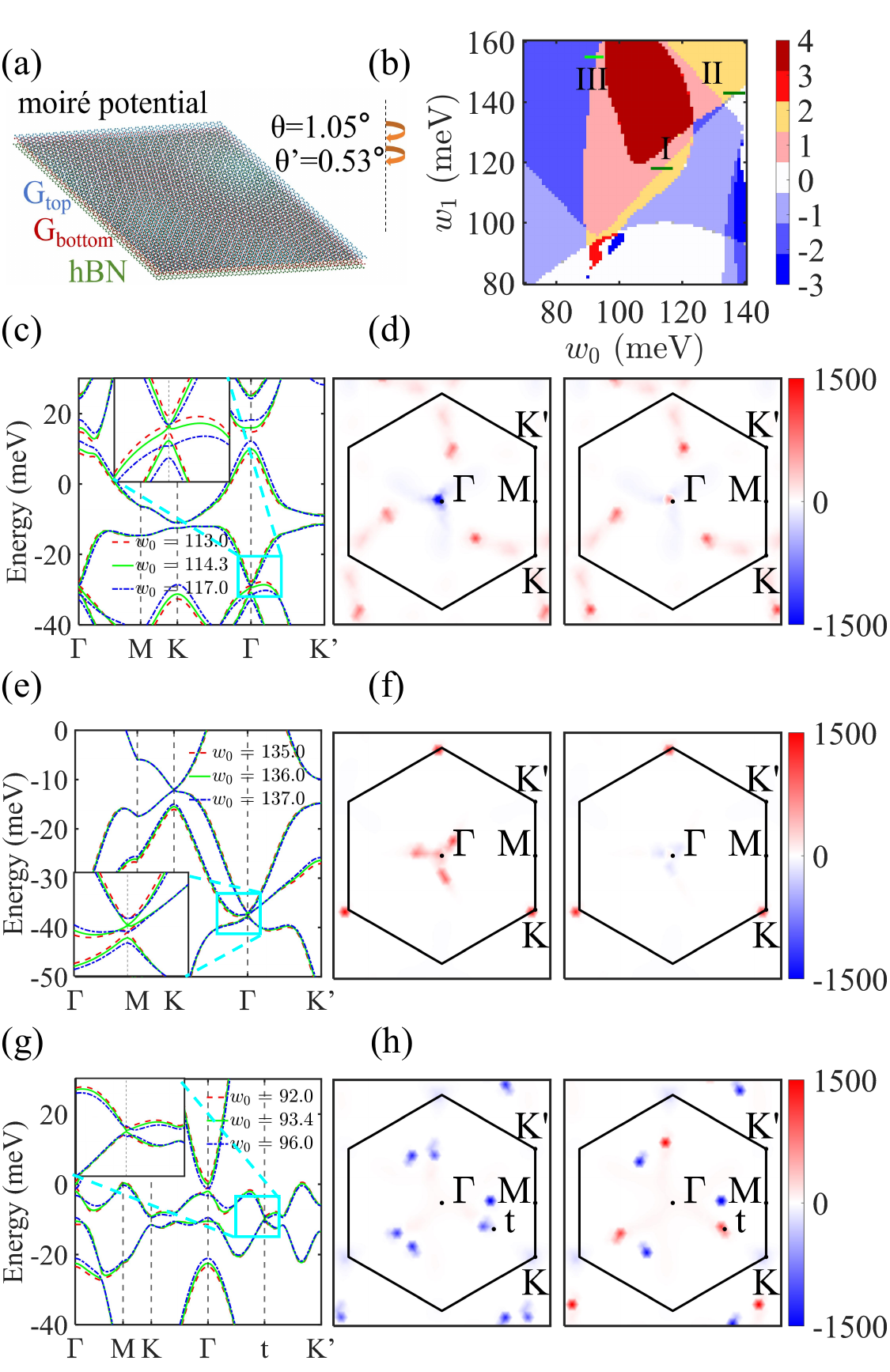}
    \caption{\label{fig:w0-w1 with moire}Topological phase transitions driven by interlayer coupling with a single aligned hBN layer. (a) Atomic schematic of TBG aligned with an hBN substrate. The twist angles $\theta$ = 1.05\degree and $\theta^\prime$ = 0.53\degree. (b) Corresponding Chern number phase diagram for the first valence band. (c) Band structure for transition I ($w_0$ = 114.3 meV, $w_1$ = 118 meV). (d) Berry curvature before (left: $w_0$ = 114 meV, C = 1) and after (right: $w_0$ = 115 meV, C = 2) transition I. (e) Band structure for transition II ($w_0$ = 136 meV, $w_1$ = 142 meV). (f) Berry curvature before (left: $w_0$ = 135 meV, C = 2) and after (right: $w_0$ = 137 meV, C = 0) transition II. (g) Band structure near the inversion point for transition III ($w_0$ = 93.4 meV,$w_1$ = 155 meV). (h) Berry curvature before (left: $w_0$ = 92 meV, C = $-2$) and after (right: $w_0$ = 94 meV, C = 1) transition III. Insets show the zoom-in band inversion point.}
\end{figure}

When the Moir\'{e} patterns from hBN-graphene and TBG interfaces overlap, a strong additional Moir\'{e} potential emerges. This potential becomes more pronounced as the twisted angle between hBN and graphene decreases. To investigate its impact on topology, we consider the configuration where the two patterns are aligned, as shown in Fig.~\ref{fig:w0-w1 with moire}(a), with the corresponding angle $\theta^{\prime}$ determined by Eq.~\ref{eq:theta and theta'}. We focus on the representative commensurate case ($\theta$ = 1.05$^{\circ}$, $\theta^\prime$ = 0.53$^{\circ}$), which exhibits a full spectrum of topological phase transitions. (A phase diagram for the other aligned configuration, ($\theta$ =1.06$^{\circ}$, $\theta^{\circ}=-0.55^{\circ}$), is provided in Appendix~\ref{appendix:B}, Fig.~\ref{fig:S3 type-2 w0-w1}.)

Figure~\ref{fig:w0-w1 with moire}(b) displays the resulting Chern number phase diagram for the first valence band. In stark contrast to the diagram without the moir\'{e} potential [Fig.~\ref{fig:w0-w1 without moire}(b)], this landscape is significantly richer, hosting a wider variety of Chern numbers (4, $\pm3, \pm2, \pm1, 0$). This enrichment directly stems from the moir\'{e} potential, which introduces additional symmetry constraints and modifies the band structure, thereby stabilizing more distinct topological phases that should be accessible in experiment. From this complex diagram, we identify multiple phase transitions characterized by changes in Chern number ($\Delta C$) of 1, 2, and 3. To understand their origins, we analyze three representative transitions labeled I, II, and III in Fig. \ref{fig:w0-w1 with moire}(b), corresponding to ($\Delta C$) of 1, 2, and 3, respectively.

Transition I ($\Delta C = 1$): This transition is associated with a band crossing at the high symmetry point such as $\Gamma$ point. Figure \ref{fig:w0-w1 with moire}(c) illustrates this for $w_0 = 114.3$ meV, $w_1$ = 118 meV, where the gap closes precisely at $\Gamma$. The Berry curvature distribution before and after the crossing [Fig. \ref{fig:w0-w1 with moire}(d)] shows a localized sign change at this high-symmetry point, accounting for a $\Delta C = \pm1$.

Transition II ($\Delta C = 2$): A distinct mechanism is observed for transitions where the Chern number changes by 2, such as 0 $\leftrightarrow$ 2, which occurs at $\Gamma$ point for $w_0 =$ 136 meV and $w_1 =$ 142 meV. As illustrated in Fig. \ref{fig:w0-w1 with moire}(e), this involves a parabolic band touching point in the BZ. The corresponding Berry curvature evolution [Fig. \ref{fig:w0-w1 with moire}(f)] confirms the transfer of two units of Chern number, differentiating it from the Dirac-point-mediated transitions for $\Delta C = 2$.

Transition III ($\Delta C = 3$): This transition is driven by band inversions at three $C_3$-symmetric points. As shown in Fig. \ref{fig:w0-w1 with moire}(g) for parameters $w_0=93.4$ meV, $w_1$ = 155 meV, a band crossing occurs at a generic t point near the $\Gamma$-K line. The corresponding reversal in Berry curvature sign at these points [Fig. \ref{fig:w0-w1 with moire}(h)] confirms the transfer of a total topological charge of 3.

\begin{figure}[h]
    \centering
    \includegraphics[width=\linewidth]{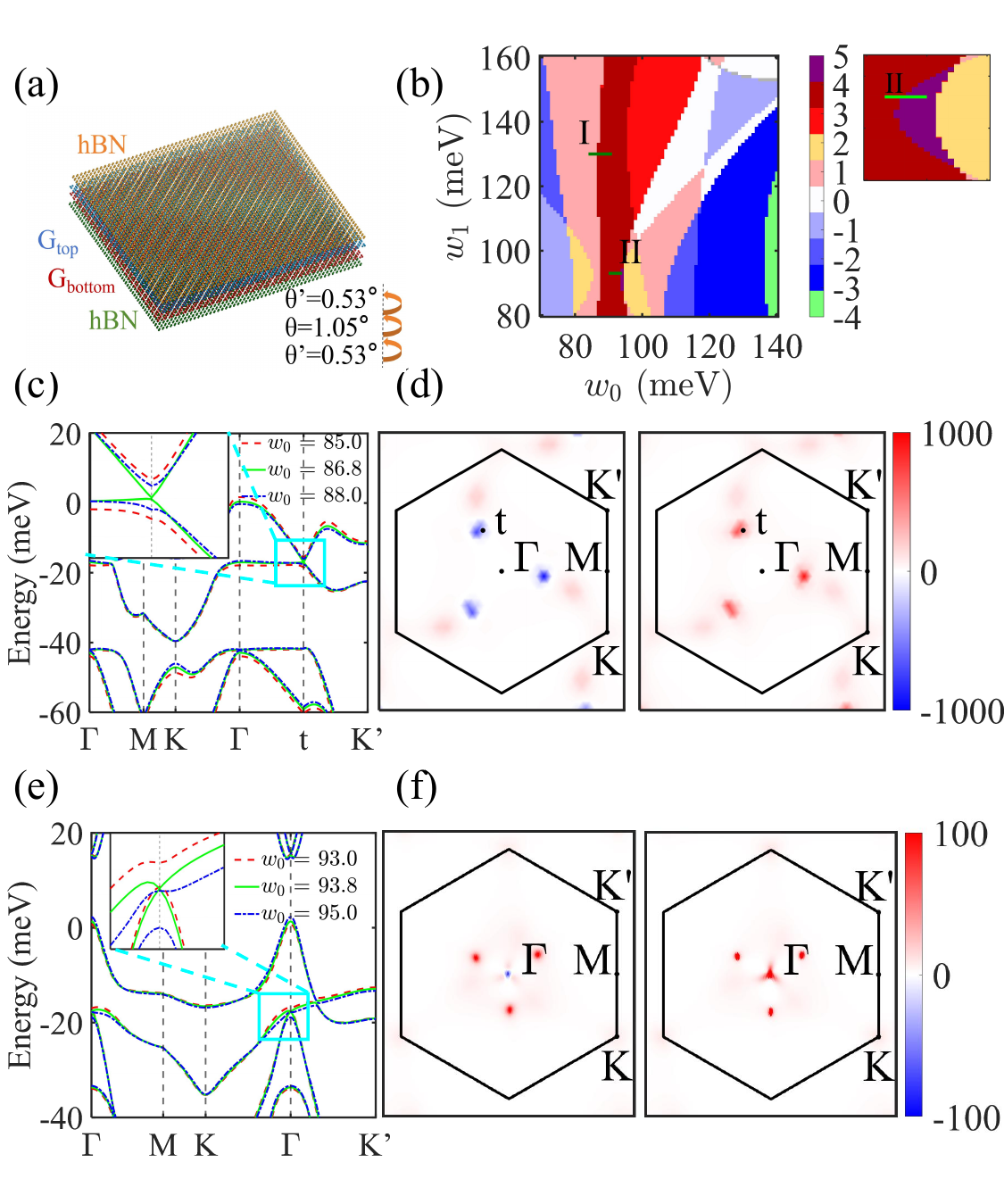}
    \caption{\label{fig:w0-w1 2L moire}Topological phase transition driven by interlayer coupling with both aligned hBN layers. (a) Atomic schematic of TBG aligned with two hBN substrates. (b) Chern number phase diagram for the first valence band. The inset in the upper right corner shows a zoom-in view of region II. 
    (c) Band structure near the band inversion point for transition I ($w_0$ = 86.8 meV, $w_1$ = 130 meV). (d) Berry curvature before (left: $w_0$ = 86 meV, C = 1) and after (right: $w_0$ = 88 meV, C = 4) transition I. (e) Band structure near the inversion point for transition II ($w_0$ = 93.8 meV, $w_1$ = 88 meV). (f) Berry curvature before (left: $w_0$ = 93 meV, C = 4) and after (right: $w_0$ = 94 meV, C = 5) transition II.}
\end{figure}

We next consider the case where TBG is encapsulated between two aligned hBN layers, as illustrated in Fig.~\ref{fig:w0-w1 2L moire}(a).  In this symmetric configuration, the moir\'{e} patterns from both the top and bottom hBN-graphene interfaces are perfectly aligned. The resulting Chern number phase diagram for the first valence band, shown in Fig.~\ref{fig:w0-w1 2L moire}(b), is substantially richer than the single-hBN case [Fig.~\ref{fig:w0-w1 with moire}(b)], featuring the emergence of two new topological phases with Chern numbers $C = -4$ and $C = 5$. 

To elucidate the mechanism driving transitions in this expanded phase space, we examine band structures and Berry curvature distributions at representative points. For transition I with $\Delta C = 3$, a band crossing occurs at a generic $C_3$ symmetric t point near the $\Gamma$-M line for parameter $w_0 = 86.8$ meV, $w_1$ = 130 meV, as shown in Fig.~\ref{fig:w0-w1 2L moire}(c). The associated reversal in the sign of the Berry curvature across this point [Fig.~\ref{fig:w0-w1 2L moire}(d)] confirms the transfer of three units of topological charge. For transition II [inset of Fig.~\ref{fig:w0-w1 2L moire}(b)] between $C = 4$ and $C = 5$ with $\Delta C$ = 1, a separate transition is mediated by a band crossing at the $\Gamma$ point for $w_0 = 93.8$ meV, $w_1$ = 88 meV as shown in Fig.~\ref{fig:w0-w1 2L moire}(e). The corresponding change in the Berry curvature distribution [Fig.~\ref{fig:w0-w1 2L moire}(f)] accounts for a change in Chern number of $\pm 1$.

In summary, the topological landscape evolves progressively with the introduction of hBN-induced potentials. Without a moir\'{e} potential, the $w_0$ and $w_1$ phase diagram already exhibits rich transitions, including phases with $C = \pm 3$. Introducing the moir\'{e} potential from a single hBN substrate further enriches the diagram, stabilizing a phase with $C = 4$. Finally, with aligned hBN on both sides, the phase space becomes most complex, giving rise to previously unobserved phases with $C = -4$ and $C = 5$. These results systematically demonstrate how the interplay between interlayer hopping and hBN-induced moir\'{e} potentials can dramatically diversify the topological phases in TBG.

\subsection{$w_0$-$\Delta_M$ topological phase diagram}
\label{2}

\begin{figure}[h]
    \centering
    \includegraphics[width=\linewidth]{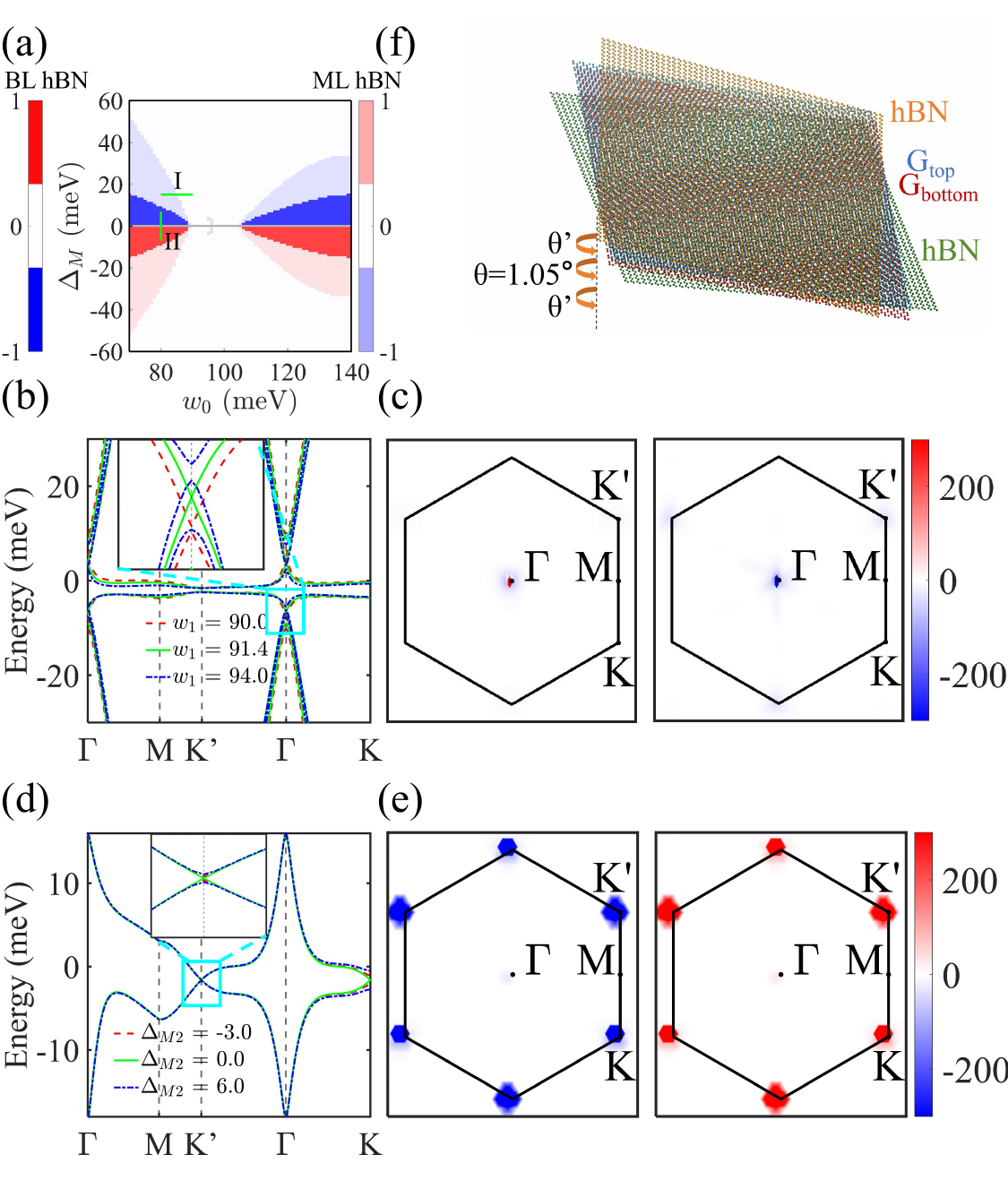}
    \caption{\label{fig:w0-deltam}Topological phase transition driven by interlayer coupling and staggered potential. (a) Chern number phase diagram for the first valence band under the condition $w_0=w_1$. Gray line mark regimes where Chern number is ill-defined. The light-colored region corresponds to TBG aligned with a single hBN substrate [Fig.~\ref{fig:w0-w1 without moire}(a)], while the dark-colored region represents TBG with hBN on both graphene layers, as shown in (f). 
    (b) Band structures near the band inversion point for transition I ($w_1$ = 91.4 meV, $\Delta_M$ = 15 meV). (c) Berry curvature before (left: $w_0=89$ meV, C = $-1$) and after (right: $w_0$ = 93 meV, C = 0) transition I. (d) Band structures near the band inversion point for transition II ($w_0$ = 80 meV, $\Delta_M$ = 0). (e) Berry curvature before ($\Delta_M$ = 0.3 meV, C = $-1$) and after ($\Delta_M$ = $-$0.3 meV, C = 1) transition II.}
\end{figure}

Having explored the role of interlayer coupling and moir\'{e} potential, we now turn to the next key parameter: the staggered potential $\Delta_M$ induced by the hBN substrate, and its interplay with interlayer hopping. Figure~\ref{fig:w0-deltam}(a) presents the topological phase diagram for the first valence band in the plane of staggered potential $\Delta_M$ versus interlayer hopping $w_0$, under the condition $w_0$ = $w_1$ (i.e., neglecting lattice relaxation). The gray line highlights the $\Delta_M$ = 0 scenario where the combined $C_{2z}T$ symmetry protects band degeneracy and the Chern number is ill-defined. The phase diagram is symmetric about the $w_0$-axis due to the $C_{2z}$ symmetry, which relates configurations with $\pm\Delta_M$ and results in opposite Chern numbers. Two structural configurations are compared. First, the light-colored region corresponds to a single hBN substrate, as shown in Fig.~\ref{fig:w0-deltam}(a), and features a wide region with Chern number $C = \pm 1$. Second, the dark-colored region represents the case with dual hBN substrates [structure shown in Fig. \ref{fig:w0-deltam}(f)], where $\Delta_{M1}=\Delta_{M2}=\Delta_{M}$. Here, the region hosting $C = \pm 1$ is noticeably narrower than in the single-substrate case. This reduction occurs because the staggered potential on both layers cooperatively enhances the band gap, promoting band inversion at smaller values of $\lvert \Delta_M \rvert$.


The topological phase transition can be understood from symmetry considerations. Without $\Delta_M$, the system preserves a combined $C_{2z}\mathcal{T}$ symmetry, which enforces degeneracy at the $K$ and $K'$ points between two flat bands, resulting in a Dirac semimetal phase. Introducing $\Delta_M$ breaks this symmetry, lifts the degeneracy and opens a gap, yielding Chern numbers of $\pm1$. As $|\Delta_M|$ increases further, the gap between the flat and dispersive bands (second valence/conduction band) narrows and eventually closes at a critical value, driving a topological phase transition from C = $\pm$1 to 0. On the other hand, interlayer coupling $w_0$ can also modify the gap between flat and dispersive bands. As $w_0$ increases, this gap initially decreases and closes, leading to a transition from C = $\pm 1$ to C = 0 due to $\Gamma$ point band inversion. With a further increase of $w_0$, the band gap subsequently increases, then decreases until it closes, resulting in a second topological phase transition from C = 0 back to C = $\pm 1$ at $\Gamma$ point.

To verify this analysis, we examine two representative transitions in Fig.~\ref{fig:w0-deltam}(a). Transition I involves a Chern number change of 1 driven by increasing $w_0$, where the first and second valence bands cross at the $\Gamma$ point. This can be seen from the evolution of the band structure [Fig.~\ref{fig:w0-deltam}(b)] and the Berry curvature distribution [Fig.~\ref{fig:w0-deltam}(c)] for $w_0 = 89$ and $93$ meV at fixed $\Delta_M = 15$ meV, showing a sign reversal across the band inversion. Transition II involves a change from $-1$ to 1, mediated by a Dirac semimetal phase as $\Delta_M$ varies, with simultaneous band crossings at $K$ and $K'$, as shown in Fig.~\ref{fig:w0-deltam}(d). The Berry curvature for $w_0=80$ meV at $\Delta_M=-$ 0.3 meV, shown in Fig.~\ref{fig:w0-deltam}(e), demonstrates opposite contributions on either side of $\Delta_M$ = 0, further validating the topological transition.

\subsection{$\Delta_{M1}-\Delta_{M2}$ topological phase diagram}
\label{3}

\begin{figure}[h]
    \centering
    \includegraphics[width=\linewidth]{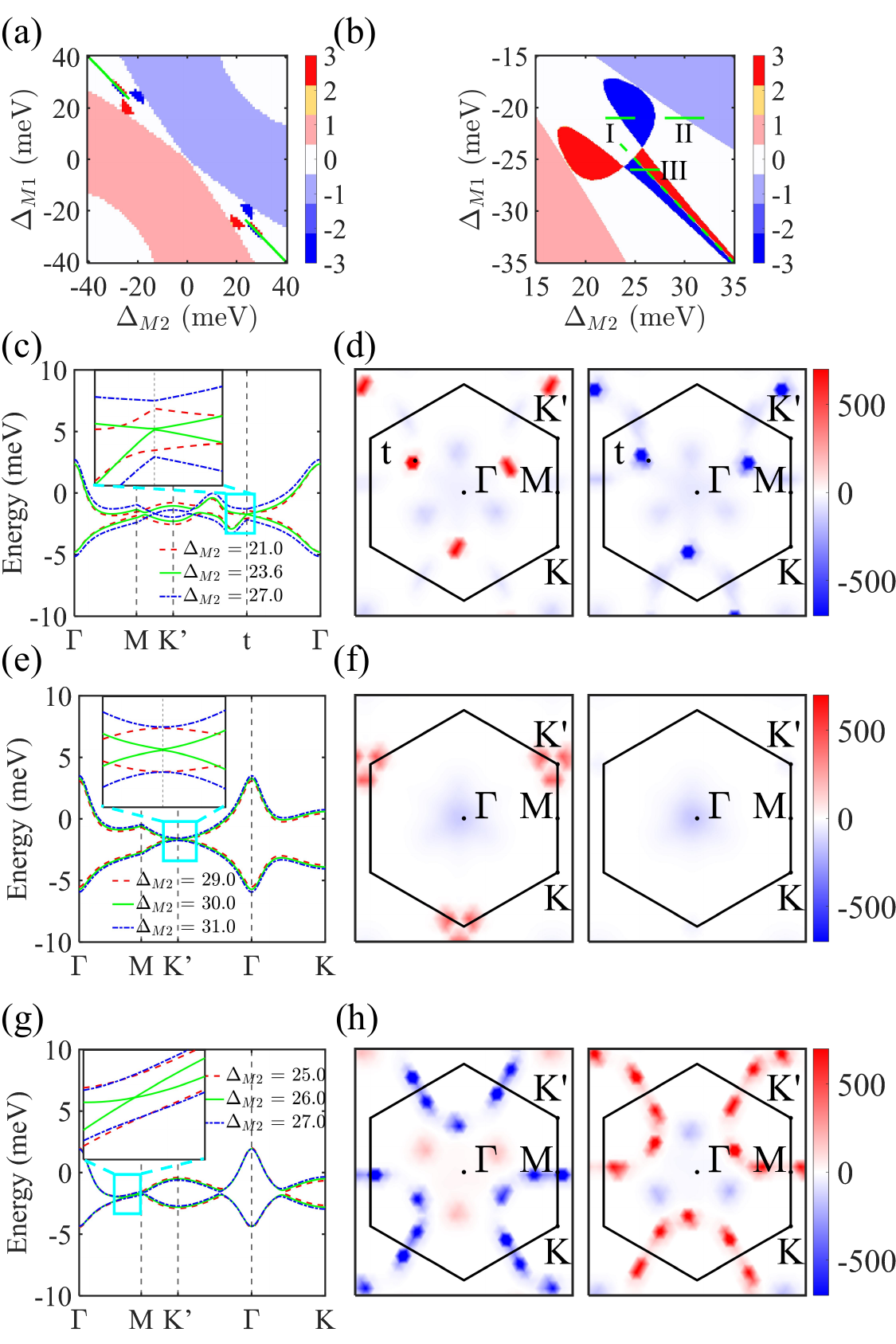}
    \caption{\label{fig:deltam12}Topological phase transition driven by staggered} potential with two misaligned hBN layers. (a) Chern number phase diagram for the first valence band. (b) Zoom-in view of (a). Along the green diagonal line, Chern number is ill-defined.
    (c) Band structures near band inversion points for transition I ($\Delta_{M2}=23.6$ meV, $\Delta_{M1}=-21$ meV). (d) Berry curvature before (left: $\Delta_{M2}=23$ meV, C = 0) and after (right: $\Delta_{M2}=24$ meV, C = $-$3) transition I. (e) Band structures near band inversion points for transition II ($\Delta_{M2}=30$ meV, $\Delta_{M1}=-21$ meV). (f) Berry curvature before ($\Delta_{M2}=29$ meV, C = 0) and after ($\Delta_{M2}=31$ meV, C = $-1$) transition II. (g) Band structures near band inversion points for transition III ($\Delta_{M2}=26$ meV, $\Delta_{M1}=-26$ meV). (h) Berry curvature before ($\Delta_{M2}=25$ meV, C = $-3$) and after ($\Delta_{M2}=27$ meV, C = 3) transition III.
\end{figure}

\begin{figure}
    \centering
    \includegraphics[width=\linewidth]{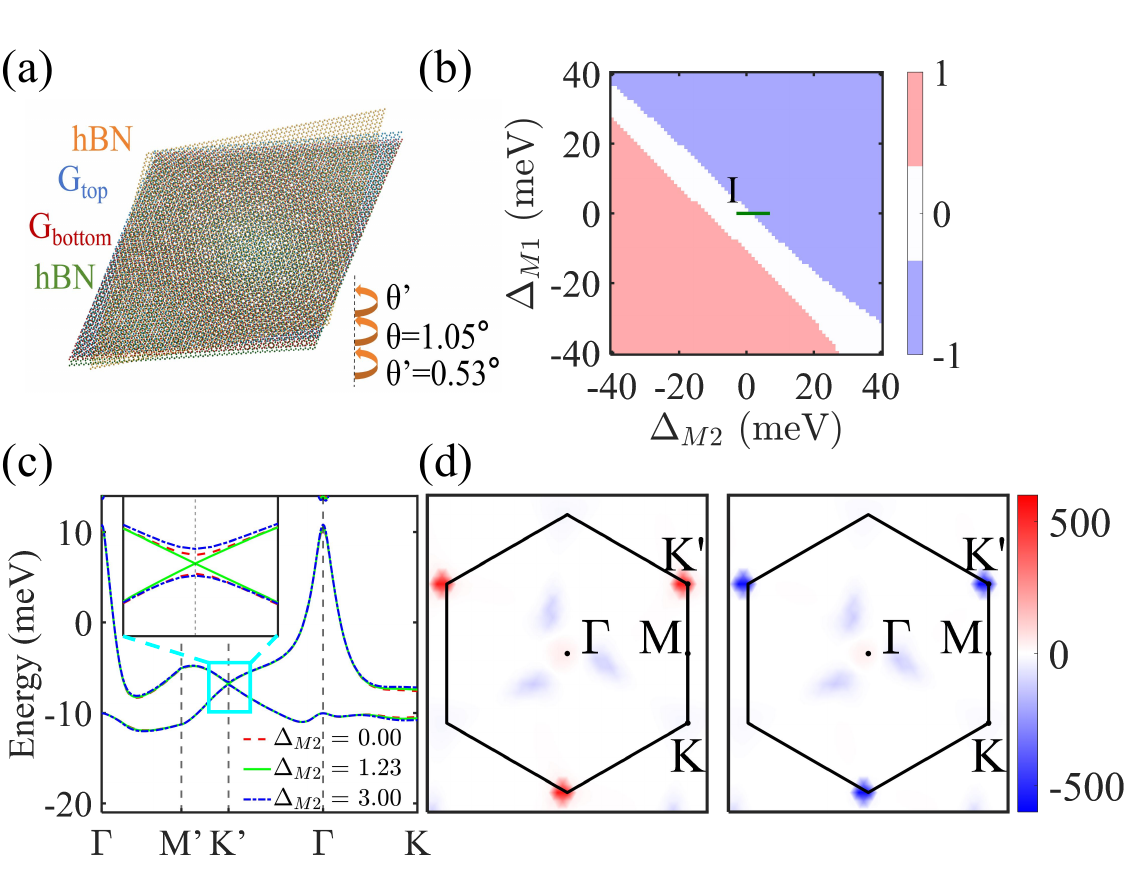}
    \caption{\label{fig:deltam12 1L moire}Topological phase transition driven by staggered potential with one layer aligned hBN and the other layer misaligned hBN. (a) Atomic schematic. (b) Chern number phase diagram for the first valence band. (c) Band structures near band inversion points for transition I ($\Delta_{M1}=0$, $\Delta_{M2}=1.23$ meV). (d) Berry curvature before ($\Delta_{M2}=0$, C = 0) and after ($\Delta_{M2}=2$ meV, C = $-1$) transition I.}
\end{figure}

Building on our analysis of symmetric staggered potentials, we now consider the more general scenario where the top and bottom hBN layers induce different staggered potentials, $\Delta_{M1}$ and $\Delta_{M2}$, on the TBG. To isolate their effect, we first analyze the case without the moir\'{e} potential, using representative interlayer hoppings $w_0 =$ 80 meV and $w_1$ = 98 meV \cite{liuOrbitalMagneticStates2021}. The resulting Chern number phase diagram for the first valence band in the $\Delta_{M1}$-$\Delta_{M2}$ plane is shown in Fig.~\ref{fig:deltam12}(a) for the dual-hBN structure in Fig.~\ref{fig:w0-deltam}(f). Beyond the expected extensive regions with C = $\pm$1, the diagram reveals compact domains with Chern numbers C = $\pm$3, which were not anticipated from simple symmetry considerations \cite{PhysRevB.103.075122, bultinckMechanismAnomalousHall2020, maoQuasiperiodicityBandTopology2021, zhangNearlyFlatChern2019}. The phase diagram is symmetric with respect to the line $\Delta_{M1}=-\Delta_{M2}$, highlighted by a green dashed line in Fig.~\ref{fig:deltam12}(a). Along this line, an effective inversion symmetry protects the degeneracy of the two flat bands, resulting in a Dirac semimetal state. Noticeably, with the change the $\Delta_M$ values, the band crossing point continuously shifts between $M$ and $\Gamma$ points, as shown in Appendix~\ref{appendix:B}, Fig.~\ref{fig:S2 ill-defined band}. Crossing this symmetry line reverses the sign of the Berry curvature, exchanging the Chern numbers +1 $\leftrightarrow$ $-$1 and +3 $\leftrightarrow$ $-$3. 

To understand the origin of these topological phases, we performed a detailed calculation of the high Chern number region, as shown in Fig.~\ref{fig:deltam12}(b). Five distinct phases are clearly resolved: $C=0$, $C=\pm$3 and $C=\pm 1$. We then analyze three characteristic transitions labeled \uppercase\expandafter{\romannumeral1}, \uppercase\expandafter{\romannumeral2}, and \uppercase\expandafter{\romannumeral3}. In the transition \uppercase\expandafter{\romannumeral1} ($C=0$ to $C=\pm3$), with $\Delta_{M1}=-21$ meV fixed, as $\Delta_{M2}$ increases from 23 meV to 24 meV, a band inversion occurs at a point along the K$^\prime$-$\Gamma$ path between the two flat bands [Fig.~\ref{fig:deltam12}(c)]. The associated sign reversal of the Berry curvature at three $C_3$-symmetric points [Fig.~\ref{fig:deltam12}(d)] leads to a change in Chern number of 3.

In the transition \uppercase\expandafter{\romannumeral2} ($C=0$ to $C=\pm1$), the transition is mediated by a band inversion at the $K'$ point [Fig.~\ref{fig:deltam12}(e)]. For parameters $\Delta_{M1}=-21$ meV and $\Delta_{M2}=29$ meV [left side of Fig.~\ref{fig:deltam12}(f)], the Berry curvature contributions from $\Gamma$ and $K^{\prime}$ points are opposite and cancel, yielding a total Chern number of zero. As $\Delta_{M2}$ increases to 31 meV [right panel of Fig.~\ref{fig:deltam12}(f)], the minimal band gap shifts to the $K^{\prime}$ point, where a band inversion occurs. After this inversion, the $\Gamma$ point develops a nonzero Berry curvature contribution, resulting in C = $-1$.

In the transition \uppercase\expandafter{\romannumeral3} ($C=-3$ to $C=3$), with $\Delta_{M1}=-26$ meV fixed, increasing $\Delta_{M2}$ from 25 meV to 27 meV drives the Chern number from $-3$ to $3$, accompanied by band inversions at six points along the $\Gamma$-M line [Fig.~\ref{fig:deltam12}(g)]. At the symmetric point $\Delta_{M2}=-\Delta_{M1}$ = 26 meV on the green dashed line, the system is a Dirac semimetal. Moving away from this line opens a gap near the $M$ point [Fig.~\ref{fig:deltam12}(h)], stabilizing phases with Chern number of $\pm3$. 


Finally, we consider the case where only one hBN layer is perfectly aligned with graphene, thereby inducing a moir\'{e} potential (Eq.~\ref{eq:moire potential}) on one graphene layer, as depicted in Fig.~\ref{fig:deltam12 1L moire}(a). The corresponding phase diagram [Fig.~\ref{fig:deltam12 1L moire}(b)] is significantly simpler, exhibiting only transitions between C = 1, 0, and $-$1. To illustrate the mechanism, we fix $\Delta_{M1}$ = 0 and vary $\Delta_{M2}$ from $-$40 meV to 40 meV. Starting from $C=1$ at $\Delta_{M2}=-$40 meV, increasing $\Delta_{M2}$ causes a band crossing at the K point, driving a transition to $C=0$. A further increase leads to a gap closure and reopening at the K$^\prime$ point, resulting in  $C=-1$. This sequence is evidenced by a clear band crossing at $\Delta_{M2}$ = 1.23 meV [Fig.~\ref{fig:deltam12 1L moire}(c)] and a sign reversal in the Berry curvature distribution [Fig.~\ref{fig:deltam12 1L moire}(d)].



\section{Summary and discussion}
\label{Summary and discussion}
In this work, we have performed a systematic theoretical study of topological phase transitions in TBG aligned with hBN. By employing the continuum model, we computed comprehensive Chern number phase diagrams across physically relevant parameter windows, focusing on the interplay between three key experimental knobs: the interlayer hopping strengths ($w_0, w_1$), the substrate-induced staggered potential ($\Delta_M$), and the moir\'{e} potential arising from lattice alignment.

Our analysis reveals a progressively richer topological landscape as the system complexity increases. In the absence of a moir\'{e} potential, variations in $w_0$ and $w_1$ alone can drive transitions to phases with Chern numbers $C=\pm3$. Introducing a moir\'{e} potential from a single aligned hBN substrate significantly expands the phase diagram, stabilizing a region with $C=4$. When TBG is symmetrically encapsulated between two hBN layers, the topological diversity is further enhanced, yielding phases with $C=-4$ and $C=5$. Furthermore, by independently tuning the staggered potentials $\Delta_{M1}$ and $\Delta_{M2}$ on the top and bottom layers, we identified compact, previously unreported domains with $C=\pm3$. Throughout this parameter space, the transitions are driven by distinct band inversion mechanisms—at generic $C_3$-symmetric points, high-symmetry points ($K$, $K'$, $\Gamma$, $M$), or through parabolic touchings—each imprinting a characteristic signature on the Berry curvature.


The phase diagrams and transition mechanisms presented here are closely connected to experiments. The parameters we have scanned ($w_0$, $w_1$, and $\Delta_M$) can vary under different physical conditions: interlayer hopping can be modified via hydrostatic pressure or dielectric environment; the staggered potential can be controlled by a perpendicular electric field; and moir\'{e} potential strength can be enhanced by pressure-driven lattice deformation. Our maps thus provide a systematic understanding of how such variations influence the topological phase transitions, clarifying how different experimental conditions (e.g., varying substrate alignment, applied gate voltages, or strain) are associated with the emergence of distinct topological phases, including the high-Chern-number phases we predict. To quantitatively assess the robustness of the predicted topological phases, we examine the band gap above and below the first valence band, bandwidth of the valence band, and gap-to-bandwidth ratio for each Chern number regime. The results are summarized in Fig.~\ref{fig:28figs}.

Looking beyond TBG, the framework and insights developed here are readily transferable to other twisted van der Waals systems where flat bands, substrate interactions, and interlayer coupling play a defining role. For instance, in twisted multilayer graphene (e.g., rhombohedral pentalayer or hexalayer graphene on hBN), the increased number of bands and interlayer degrees of freedom, combined with substrate effects, could host an even wider array of topological and correlated phases. The recent experimental discovery of fractional Chern insulators in such systems underscores their potential. Similarly, in twisted transition metal dichalcogenides (e.g., MoTe$_2$), where strong light-matter coupling and spin-valley physics are at play, the introduction of a tunable moir\'{e} potential from an encapsulating layer could enable new pathways to control topology and many-body states.

\acknowledgments
The authors thank Junxi Yu for helpful discussions. This work is supported by the National Key R\&D Program of China (No. 2024YFA1408400), the National Natural Science Foundation of China (No. 12204037) and the Beijing Institute of Technology Research Fund Program for Young Scholars.

\appendix

\section{Details of Hamiltonian}
\label{appendix:BM model}

The electronic structures of TBG system can be well described by the continuum model~\cite{bistritzerMoireBandsTwisted2011}, which is also called Bistritzer-MacDonald (BM) model.
The low energy states of graphene are dominated by $p_z$ orbital near the $K$ and $K^{\prime}$ points.
For aligned bilayer graphene ($\theta=0$), the $k\cdot p$ Hamiltonian describing the displacement dependence can be expressed as
\begin{equation}
    H_G(\theta=0,\boldsymbol{d_0})=\begin{pmatrix} -\hbar v_f\boldsymbol{q}\cdot\sigma^{\mu} & T(\boldsymbol{d_0}) \\ T^\dagger(\boldsymbol{d_0}) & -\hbar v_f\boldsymbol{q}\cdot\sigma^{\mu} \end{pmatrix},
\end{equation}
with
\begin{align}
    T(\boldsymbol{d}_0)=&\begin{pmatrix} w_0 & w_1 \\ w_1 & w_0 \end{pmatrix} + e^{i\mu\boldsymbol{b}_2 \cdot \boldsymbol{d}_0} \begin{pmatrix} w_0 e^{-i\mu \phi} & w_1 \\ w_1 e^{i\mu \phi} & w_0 e^{-i\mu \phi} \end{pmatrix} \notag \\
    &+ e^{-i\mu\boldsymbol{b}_1 \cdot \boldsymbol{d}_0} \begin{pmatrix} w_0 e^{i\mu\phi} & w_1 \\ w_1 e^{-i\mu\phi} & w_0 e^{i\mu\phi} \end{pmatrix},
\end{align}
where $\sigma^\mu=(\mu\sigma_x,\sigma_y)$ with $\sigma_x, \sigma_y$ being Pauli matrices, and $\mu=\pm1$ represents the valley ($K^\prime$ for $\mu=1$, $K$ for $\mu=-1$). The reciprocal lattice vectors of monolayer graphene are $K=(\frac{4\pi}{3\sqrt{3}d},0), K^{\prime}=(-\frac{4\pi}{3\sqrt{3}d},0)$,
$\boldsymbol{b}_1=\frac{4\pi}{3d}(\frac{\sqrt{3}}{2},\frac{1}{2}), \boldsymbol{b}_2=\frac{4\pi}{3d}(-\frac{\sqrt{3}}{2},\frac{1}{2})$. The momentum $\boldsymbol{q}$ measures the deviation from the Dirac points of two graphene layers, so that $\boldsymbol{k} = K^\pm + \boldsymbol{q}$. The phase $\phi =\frac{2\pi}{3} $, and the lattice constant of monolayer graphene is $a=\sqrt{3}d$ with $\boldsymbol{a}_1=a(\frac{1}{2},\frac{\sqrt{3}}{2}), \boldsymbol{a}_2=a(-\frac{1}{2},\frac{\sqrt{3}}{2})$ are lattice vectors of monolayer graphene. The displacement $\boldsymbol{d}_0$ distinguishes the stacking: $\boldsymbol{d}_0=0$ for AB stacking; $\boldsymbol{d}_0=\frac{1}{3}(\boldsymbol{a_1}+\boldsymbol{a_2})$ for AA stacking; $\boldsymbol{d}_0=-\frac{1}{3}(\boldsymbol{a}_1+\boldsymbol{a}_2)$ for BA stacking. The bulk Fermi velocity is $\hbar v_f = 5.25 \text{eV} \stackrel{\circ}{A}$. The interlayer hopping strengths $w_0$ (AA/BB) and $w_1$ (AB/BA) can be tuned by varying the interlayer distance. 
The local displacement $\boldsymbol{d_0}$ can be approximated as $\boldsymbol{d}_{0}=\theta\hat{z} \times\boldsymbol{r}$ \cite{jungInitioTheoryMoire2014}. Within this approximation, the moir\'{e} Hamiltonian takes the form
\begin{equation}
    H_G(\boldsymbol{r})=\begin{pmatrix} -\hbar v_f\boldsymbol{q}\cdot\sigma^{\mu} & T(\boldsymbol{r}) \\ T^\dagger(\boldsymbol{r}) & -\hbar v_f\boldsymbol{q}\cdot\sigma^{\mu} \end{pmatrix}
    \label{BM hamiltonian},
\end{equation}
with
\begin{align}
    T(\boldsymbol{r})=&\begin{pmatrix} w_0 & w_1 \\ w_1 & w_0 \end{pmatrix} + e^{i\mu\boldsymbol{b}_{m2} \cdot \boldsymbol{r}} \begin{pmatrix} w_0 e^{-i\mu \phi} & w_1 \\ w_1 e^{i\mu \phi} & w_0 e^{-i\mu \phi} \end{pmatrix} \notag \\
    &+ e^{-i\mu{\boldsymbol{b}_{m1}} \cdot \boldsymbol{r}} \begin{pmatrix} w_0 e^{i\mu\phi} & w_1 \\ w_1 e^{-i\mu\phi} & w_0 e^{i\mu\phi} \end{pmatrix},
    \label{interlayer coupling}
\end{align}
where we use the relationship $\boldsymbol{b}_i\cdot \boldsymbol{d}_0=\boldsymbol{b}_{mi}\cdot\boldsymbol{r}$. Here $\boldsymbol{b}_{mi}$ are the moir\'{e} reciprocal lattice vectors with $\boldsymbol{b}_{m1}=\frac{4\pi}{\sqrt{3}a}(\sin\theta,1-\cos\theta)$, and $\boldsymbol{b}_{m2}=R_{2\pi/3}\boldsymbol{b_{m1}}$ with $R_{2\pi/3}=\begin{pmatrix}
-\frac{1}{2}   &-\frac{\sqrt{3} }{2}  \\
  \frac{\sqrt{3} }{2} &-\frac{1}{2} 
\end{pmatrix}$ being rotation matrix with a 120$^\circ$ degree.

For the calculations without moir\'{e} potential (Figs.~\ref{fig:w0-w1 without moire},~\ref{fig:w0-deltam}, and~\ref{fig:deltam12}), we set the moir\'{e} potential to zero and diagonalize the Hamiltonian in a plane-wave basis expanded in the reciprocal lattice vectors of the TBG moir\'{e} superlattice. For the calculations with the moir\'{e} potential (Figs.~\ref{fig:w0-w1 with moire},~\ref{fig:w0-w1 2L moire}, and~\ref{fig:deltam12 1L moire}), we included the hBN moir\'{e} potential $V(\boldsymbol{G_j'})$ from Eq.~\ref{eq:moire potential}. Since the TBG/hBN is doubly commensurate in the aligned configurations studied here, meaning the two moir\'{e} patterns share the same reciprocal lattice vectors, the combined potential is periodic with respect to the TBG moir\'{e} Brillouin zone. We therefore retain the TBG moir\'{e} Brillouin zone as the basis for the band structure calculations.

\section{Chern number and Berry curvature calculation method}
\label{appendix:numerical calaulation of chern}
Our calculation of Chern number and Berry curvature follows Ref.~\cite{haldaneModelQuantumHall1988,fukuiChernNumbersDiscretized2005a}. 
We discretize the moir\'{e} BZ with N $\times$ N small chunks. The flux on the small chunks is gauge invariant and is calculated by
\begin{equation}
    \Phi_{n}=\operatorname{Arg}\left(\prod_{p}\left\langle u\left(\mathbf{k}_{n, p}\right) \mid u\left(\mathbf{k}_{n, p+1}\right)\right\rangle\right),
\end{equation}
where n is band index and we calculate the first valence band throughout the paper, p is a corner of the small chunk and $\boldsymbol{k}_{n,p}$ is the momenta.

Chern number is calculated by 
\begin{equation}
    C=\frac{\sum_{n}\Phi_n}{2\pi} \operatorname{mod} 1=\frac{\Phi}{2\pi} \operatorname{mod} 1,
\end{equation}
where $\Phi\operatorname{mod}2\pi$ is the total flux through moir\'{e} BZ and also called Berry phase.
It is the integral of Berry curvature $\mathbf{B(\mathbf{k})}$
\begin{equation}
    \Phi=\oint_{\mathbf{k}} d^{2} \mathbf{k} \cdot \mathbf{B}(\mathbf{k}),
\end{equation}
So Berry curvature is calculated by
\begin{equation}
    \mathbf{B(\mathbf{k})}=\frac{\Phi_n}{d^{2} \mathbf{k}}.
\end{equation}

We choose N as 25 for efficiency when scanning large parameter spaces. Near the phase boundaries where the band gap becomes very small, we employ a refined k-mesh (N $=$ 100) to ensure that the Chern number is accurately resolved. The Berry curvature is reported in units of \AA$^{-2}$.
The Chern number phase diagrams presented in this work are constructed from calculations performed on a dense parameter grid, with the key parameters varied in steps of 1 meV. We use truncation equal to four under a plane wave basis. We found that it is enough to get accurate low energy band structures and Chern number.
To confirm numerical convergence, we have performed systematic tests with a 100 $\times$ 100 k-mesh and a larger truncation ($K_{\text{tr}} = 6$) for representative parameter points: $(w_0, w_1) = (100, 120)$ meV ($C = +3$) and $(120, 140)$ meV ($C = +1$) from Fig.~\ref{fig:w0-w1 without moire}(b); $(w_0, w_1) = (110, 140)$ meV ($C = +4$) from Fig.~\ref{fig:w0-w1 with moire}(b); and $(w_0, w_1) = (90, 120)$ meV ($C = +4$) from Fig.~\ref{fig:w0-w1 2L moire}(b). In all cases, the Chern numbers remain unchanged under the refined numerical parameters, confirming the robustness of the reported high-Chern-number phases.

\section{Supplementary of phase diagrams}
\label{appendix:B}

\begin{figure}
    \centering
    \includegraphics[width=0.9\linewidth]{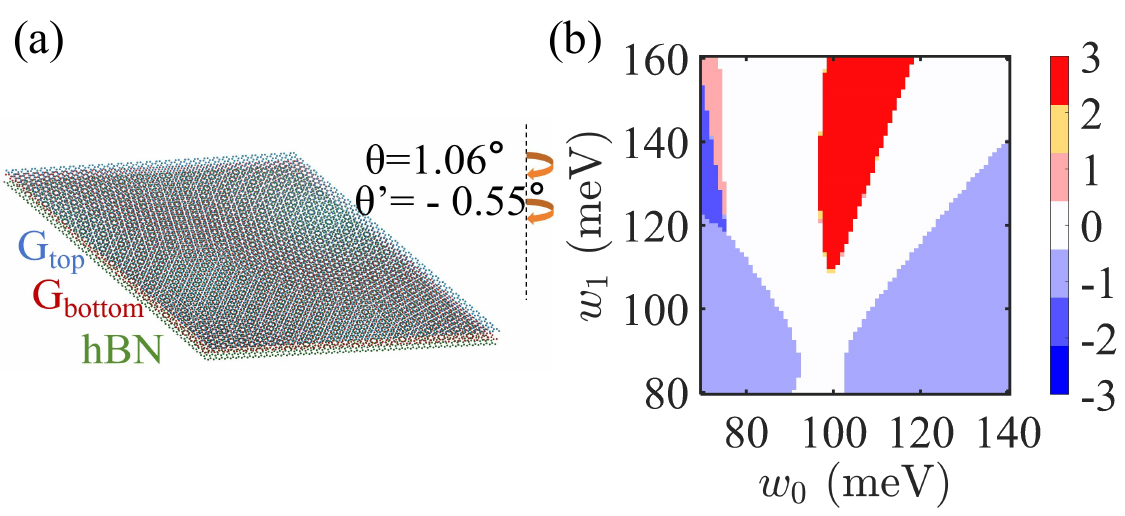}
    \caption{\label{fig:S3 type-2 w0-w1}Topological phase diagram of a commensurate TBG/hBN structure with $\theta = 1.06^\circ$ and $\theta' = -0.55^\circ$. (a) Schematics of the atomic configuration. (b) Chern number phase diagram of the first valence band in the ($w_0$, $w_1$) plane.}
\end{figure}

In the main text, we presented the topological phase diagram for a commensurate TBG/hBN structure with $\theta=1.05^\circ, \theta^\prime=0.53^\circ$. To demonstrate that the rich topological landscape is not unique to that particular geometry, we show here the phase diagram for a structurally distinct but equally commensurate configuration where the hBN substrate is rotated in the opposite sense, corresponding to $\theta = 1.06^\circ$ and $\theta' = -0.55^\circ$. Fig.~\ref{fig:S3 type-2 w0-w1}(a) illustrates the corresponding atomic arrangement. The resulting Chern‑number phase diagram in the $w_0$-$w_1$ plane, displayed in panel Fig.~\ref{fig:S3 type-2 w0-w1}(b), reveals a different pattern of topological phases and phase boundaries compared to Fig.~\ref{fig:w0-w1 with moire}(b) in the main text. This comparison underscores that while the specific sequence and Chern numbers of the phases depend sensitively on the relative stacking direction, the emergence of multiple, tunable topological phases is a generic feature of commensurate TBG/hBN moir\'{e} systems.

\begin{figure*}[!t]
    \centering
    \includegraphics[width=\linewidth]{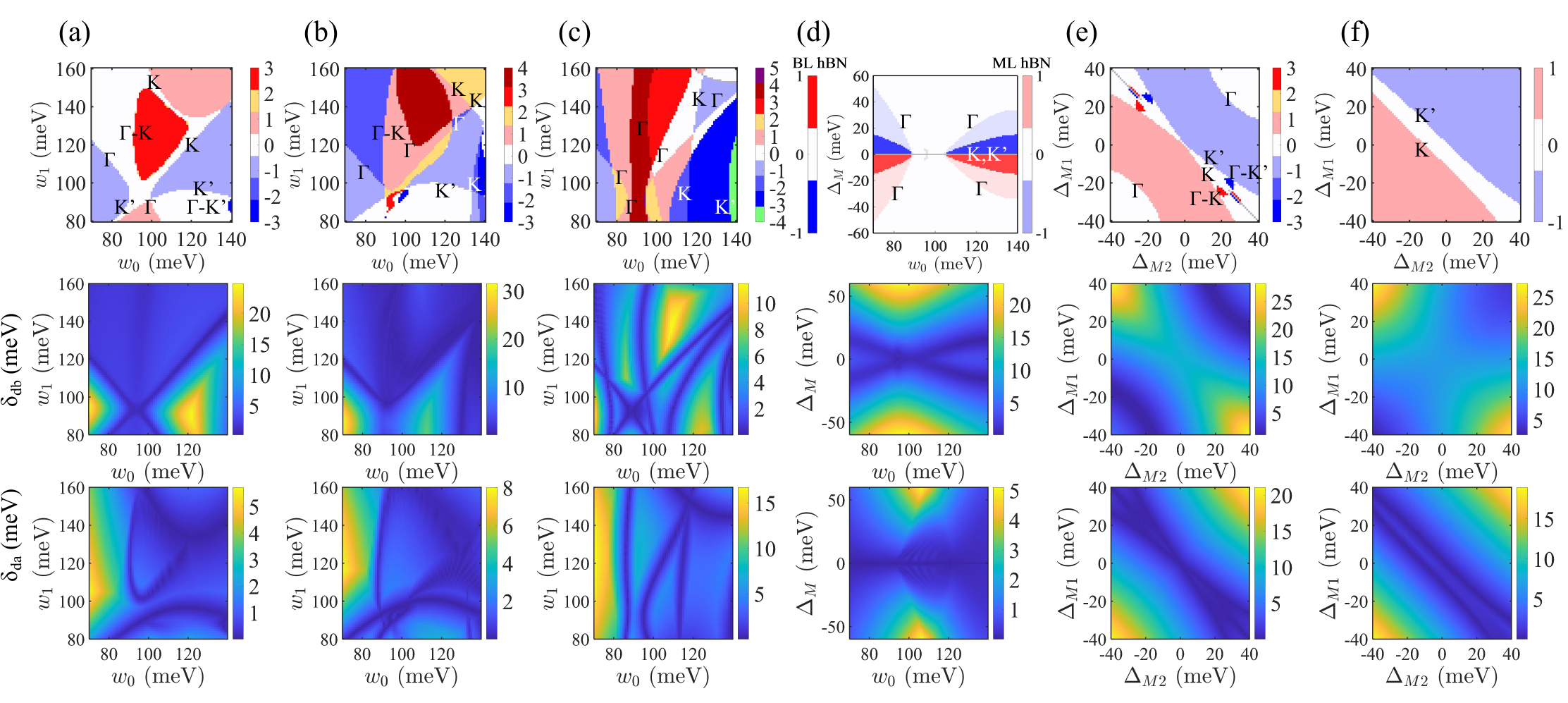}
    \caption{\label{fig:direct_gap}From top to bottom rows we show phase diagram, direct band gap below the first valence band ($\delta_{db}$), direct band gap above the first valence band ($\delta_{da}$) for (a)-(c) Figs.~\ref{fig:w0-w1 without moire}-\ref{fig:w0-w1 2L moire}, (d), dark-shaded region in Fig.~\ref{fig:w0-deltam}, and (e), (f) Figs.~\ref{fig:deltam12}-\ref{fig:deltam12 1L moire}.}
\end{figure*}

In the first row of Fig.~\ref{fig:direct_gap}, we mark all band inversion points corresponding to the phase diagrams presented in the main text. For topological phase transitions that involve a Chern number change of 3, we do not label every individual transition point because the band inversion points can shift with parameters. As an example, for the $C=1$ to 4 phase transition shown in the first row of Fig.~\ref{fig:direct_gap} the band inversion point shifts from a location along the $\Gamma$-K line to a point on the $\Gamma$-M line as $w_1$ increases. 

\begin{figure}[h]
    \centering
    \includegraphics[width=0.9\linewidth]{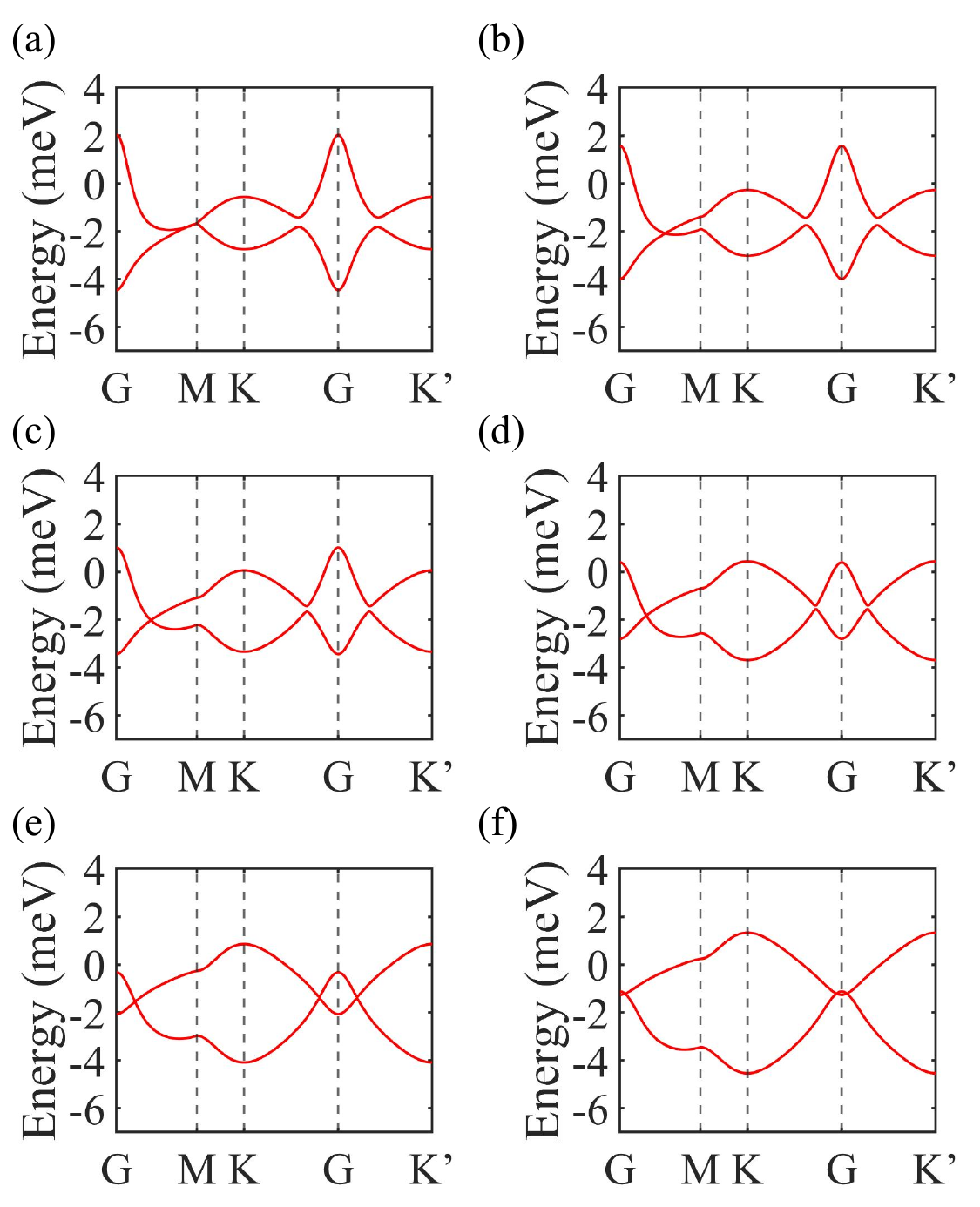}
    \caption{\label{fig:S2 ill-defined band}Band structures of $\Delta_{M1}=-\Delta_{M2}$ with $w_0$ = 80 meV, $w_1$ = 98 meV. More details are in Fig.~\ref{fig:deltam12}. (a) $\Delta_{M1}=-$25 meV, $\Delta_{M2}=$ 25 meV. (b) $\Delta_{M1}=-$30 meV, $\Delta_{M2}=$ 30 meV. (c) $\Delta_{M1}=-$35 meV, $\Delta_{M2}=$ 35 meV. (d) $\Delta_{M1}=-$40 meV, $\Delta_{M2}=$ 40 meV. (e) $\Delta_{M1}=-$45 meV, $\Delta_{M2}=$ 45 meV. (f) $\Delta_{M1}=-$50 meV, $\Delta_{M2}=$ 50 meV.}
\end{figure}

The evolution of the band crossing along the degenerate line $\Delta_{M1} = -\Delta_{M2}$ is depicted in Fig.~\ref{fig:S2 ill-defined band}. There, we plot the band structures for a series of representative points: $(\Delta_{M1}, \Delta_{M2}) = (-25,25)$, $(-30,30)$, $(-35,35)$, $(-40,40)$, $(-45,45)$, and $(-50,50)$ meV, all calculated at $w_0 = 80$ meV and $w_1 = 98$ meV. The sequence demonstrates a continuous shift of the crossing point from the $M$ point toward the $\Gamma$ point as $\Delta_{M1}$ increases from 24 meV to 40 meV or decreases from $-$24 meV to $-$40 meV.

\section{Band gap, bandwidth, gap-to-bandwidth ratio}

\begin{figure*}[t]
    \centering
    \includegraphics[width=\linewidth]{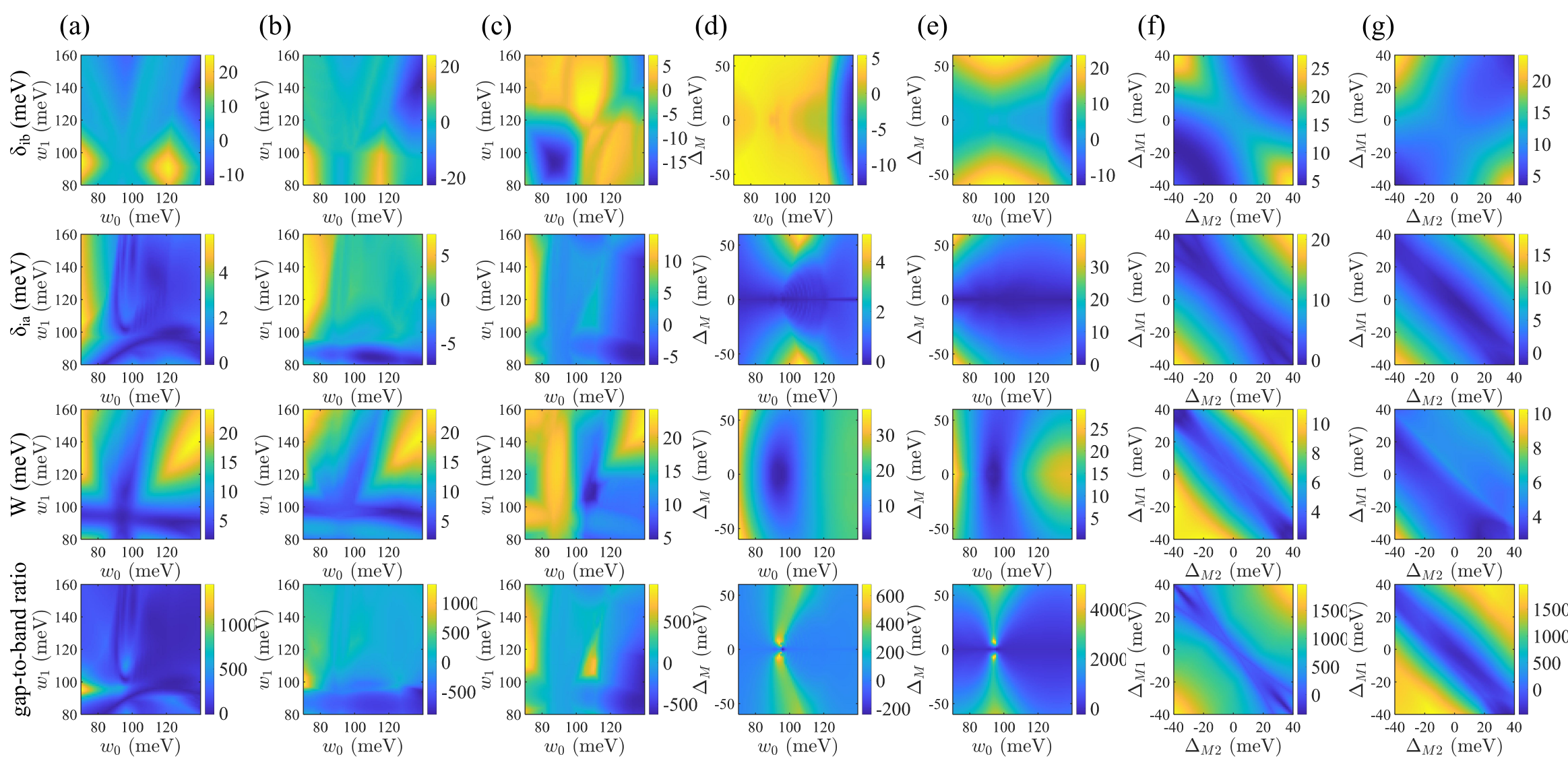}
    \caption{From top to bottom rows we show indirect band gap below the first valence band ($\delta_{ib}$), indirect band gap above the first valence band ($\delta_{ia}$), bandwidth ($W$) of the first valence band and the ratio between the gap of the first valence and conduction bands to the bandwidth of the first valence band for (a)-(c) Figs.~\ref{fig:w0-w1 without moire}-\ref{fig:w0-w1 2L moire}, (d), (e) light-shaded and dark-shaded region in Fig.~\ref{fig:w0-deltam}, and (f), (g) Figs.~\ref{fig:deltam12}-\ref{fig:deltam12 1L moire}.}
    \label{fig:28figs}
\end{figure*}

Figure~\ref{fig:28figs} show (i) the indirect gap below the first valence band, (ii) the indirect gap above the first valence band, (iii) its bandwidth, and (iv) the ratio of gap to bandwidth, respectively, for the parameter ranges studied in the main text. Notably, for each Chern number regime, there exists a finite parameter range where band gap is sufficiently large to stabilize the high-Chern-number phases. Moreover, the gap-to-bandwidth ratios in these regions are also big enough, making these phases promising platforms for investigating correlated effect in flat bands with high Chern numbers.

\FloatBarrier

%

\end{document}